\documentclass[12pt,a4paper]{article}

\usepackage{graphics}
\usepackage{latexsym}
\usepackage{amsmath}
\usepackage{amssymb}
\usepackage{slashed}
\usepackage[usenames]{color}

\title{Comparing measurements and limits on the warm dark
  matter temperature-to-mass ratio}
\author{Bruce Hoeneisen}
\date{\small{
Universidad San Francisco de Quito, Quito, Ecuador \\
Email: bhoeneisen@usfq.edu.ec \\
20 August 2023}
}

\begin{document}
\maketitle

\begin{abstract}
\noindent
Limits on the ``thermal relic" warm dark matter mass are
apparently inconsistent with measurements of the warm dark
matter temperature-to-mass ratio. We try to understand this problem.
\end{abstract}

\section{Introduction}

\textit{Limits} on the ``thermal relic" warm dark matter mass have been obtained from
the Lyman-$\alpha$ flux power spectrum of quasars and hydro-dynamical simulations.
For example, \cite{Viel1} obtains $m_x > 550$ eV at $2 \sigma$ in 2005,
\cite{Viel} obtains $m_x > 3300$ eV at $2 \sigma$ in 2013, and
\cite{Madau} obtains $m_x > 3100$ eV at 95 \% confidence in 2023.

On the other hand, \textit{measurements} of the non-relativistic warm dark matter
temperature-to-mass ratio have been obtained from the rotation curves of 10 spiral
galaxies measured by the THINGS collaboration \cite{part1} \cite{part2}, 46 spiral galaxies measured by
the SPARC collaboration \cite{sparc} \cite{three}, and 11 dwarf spiral galaxies measured by the 
LITTLE-THINGS collaboration \cite{little_things}; and \textit{independently}, from galaxy stellar mass
distributions \cite{stellar} \cite{cold} \cite{fermion}, 
galaxy rest frame ultra-violet luminosity distributions \cite{UV} 
\cite{missing_satellite},
from first galaxies \cite{stellar} \cite{first}, and from the re-ionization optical depth \cite{UV} \cite{first}.
These measurements are consistent (even tho they depend on the different detailed scenarios
assumed for each measurement).
A summary of the final results is presented in \cite{poets}. Details of the final measurements
can be found in \cite{UV} and \cite{little_things}.

It turns out that the measured warm dark matter temperature-to-mass ratio happens to
coincide with the ``no freeze-in and no freeze-out" scenario of spin zero warm dark matter
particles that decouple early on from the Standard Model sector. 
Spin one-half and spin one dark matter are disfavored because they correspond to a 
temperature-to-mass ratio larger than measured.
As an example, if spin zero
warm dark matter particles couple to the Higgs boson, their mass is $M_S \approx 150 \pm 2$ eV.
This uncertainty only includes the contribution from the measured dark matter density.
The mass depends mildly ($\approx \pm 50\%$) 
on the temperature at which the dark matter particles decouple from
the Standard Model sector, the spin of the dark matter particles,
and on the assumed momentum distribution of the non-relativistic
dark matter. The range of measurements and masses for several scenarios is summarized
in Table 3 of \cite{UV}. 

\begin{figure}
\begin{center}
\scalebox{0.7}
{\includegraphics{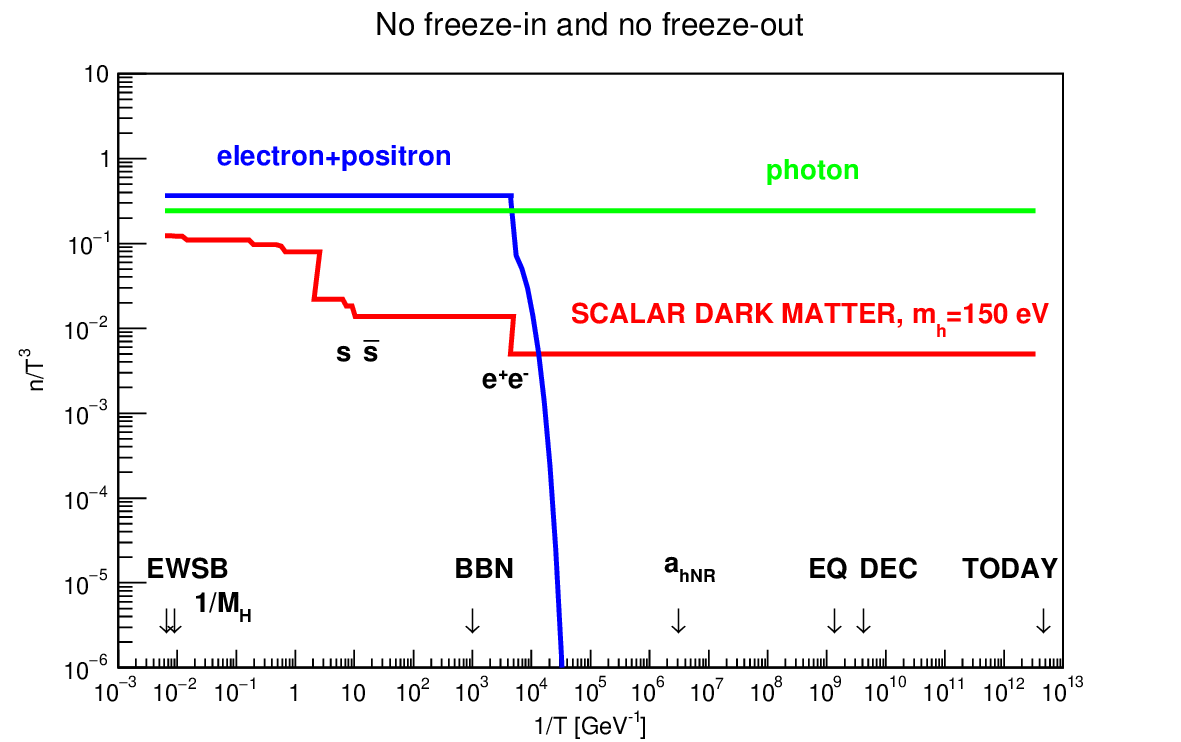}}
\caption{\small{
The ``no freeze-in and no freeze-out" warm dark matter scenario
is illustrated for spin zero warm dark matter particles coupled to the Higgs boson.
$T$ is the \textit{photon} temperature, and the $n$'s are particle number densities.
The abbreviations stand for ``Electro-Weak Symmetry Breaking", ``Big Bang Nucleosynthesis",
``EQuivalence" of matter and radiation densities, and ``DECoupling" of photons from
Standard Model matter. Dark matter particles become non-relativistic at $a_{h\textrm{NR}}$.
Time advances towards the right. From \cite{UV} \cite{poets}.
}}
\label{example_no_fio}
\end{center}
\end{figure}

Let me explain.
``No freeze-in" means that dark matter reaches thermal and diffusive, i.e. chemical,
equilibrium, in the early universe, with the particles of the
Standard Model of Quarks and Leptons.
Dark matter then decouples from the Standard Model sector while still ultra-relativistic.
``No freeze-out" means that, when dark matter becomes non-relativistic, its
self interactions are so weak that these particles do not annihilate each other.
An example of ``no freeze-in and no freeze-out" spin zero dark matter 
coupled to the Higgs boson is shown in Figure \ref{example_no_fio} \cite{poets}.

In this note we try to understand the difference between these \textit{measurements}
and \textit{limits}.
Our conclusion is that the limits on the ``thermal relic" 
warm dark matter mass do not apply to the \textit{colder}
``no freeze-in and no freeze-out" scenario of warm dark matter.

\section{Comparison}

The reason for this conclusion is as follows. If $m_x \gg 150$ eV, then dark matter
necessarily freezes-in and/or freezes-out, in order to not exceed the measured dark matter 
density, see Figure \ref{example_no_fio}. 
There are several scenarios. In	any case,
``thermal relic" dark matter has the photon temperature at the time   
of ultra-relativistic or non-relativistic decoupling from the Standard Model sector.
Therefore the ``no freeze-in and no freeze-out" warm dark matter
is \textit{colder} than a ``thermal relic", after $e^+ e^-$ recombination and 
while still ultra-relativistic,	by a factor less than, or equal	to,
$T_S/T_\gamma = (43/(11 g_\textrm{dec}))^{1/3}$ \cite{PDG2020}. This factor is $0.345$ for the
example	of dark matter coupled to the Higgs boson.
$g_\textrm{dec} = \sum{N_b} + (7/8) \sum{N_f}$ at decoupling
of ultra-relativistic ``no freeze-in and no freeze-out" 
dark matter from the Standard Model sector \cite{PDG2020}. 
$N_b$ or $N_f$ are the number of boson or fermion spin polarization's, respectively. 
Hence, to compare $M_S$ with limits $m_x$
it is necessary to multiply $m_x$ by a factor $< 0.345^{1.11} = 0.31$
(the index 1.11 is taken from equation (2) of \cite{Viel}).

As an alternative estimate,
let us translate the limit $m_x > 3300$ eV at $2 \sigma$ of \cite{Viel} into a limit on $M_S$.
From equation (2) of \cite{Viel}, i.e.
\begin{equation}
k_{1/2} \approx 6.5 \frac{h}{\textrm{Mpc}} \left( \frac{m_x}{1 \textrm{keV}} \right)^{1.11}
\left( \frac{\Omega_c}{0.25} \right)^{-0.11} \left( \frac{h}{0.7} \right)^{1.22},
\end{equation}
valid for ``thermal relic" dark matter (note that $k_{1/2}$ is really a function of $m_x/T_x$),
we obtain $k_{1/2} > 15.64$ Mpc$^{-1}$
(we take all cosmological parameters from \cite{PDG2020}).
This is the comoving wave-vector at which the warm dark matter free-streaming
reduces the cold dark matter power spectrum by a factor 2, i.e.
$\tau^2(k_{1/2}) = 1/2$, with
$\tau^2(k) \equiv P_\textrm{WDM}(k)/P_\textrm{CDM}(k)$.
The power spectrum is reduced by a factor $e$ at 
$k_{1/e} = k_{1/2}/0.83255 > 18.79$ Mpc$^{-1}$.
From Figure 4 of \cite{Boyanovsky} we obtain the comoving power spectrum cut-off at the time of
equal matter and radiation densities (before the non-linear re-generation of small
scale structure):
\begin{equation}
k_\textrm{fs}(t_\textrm{eq})
= \frac{1.455}{\sqrt{2}} \sqrt{\frac{4 \pi G \bar{\rho}_c(1) a_\textrm{eq}}{v_{h\textrm{rms}}(1)^2}}
\end{equation}
(as a cross-check, this expression agrees with a numerical integral).
$v_{h\textit{rms}}(a) = v_{h\textit{rms}}(1)/a$ is the
root-mean-square velocity of non-relativistic dark matter particles at expansion parameter $a$.
Equating $k_{1/e} = k_\textrm{fs}(t_\textrm{eq})$ we obtain
$v_{h\textit{rms}}(1) < 39.89$ m/s.
From (28) of \cite{Pfenniger} we obtain, for the no freeze-out scenario,
\begin{equation}
M_S = 108 \left( \frac{0.76 \textrm{ km/s}}{v_{h\textrm{rms}}(1)} \right)^{3/4}
\left( \frac{1}{N_b} \right)^{1/4} \textrm{ eV}.
\end{equation}
With $N_b = 1$, we finally obtain
$M_S > 985$ eV. Note that $M_S/m_x = 985/3300 = 0.30$, in good agreement
with the estimate of the previous paragraph.

The free-streaming power spectrum suppression factor relative to 
the cold dark matter scenario has the approximate form
\begin{eqnarray}
\tau^2(k) & = & \exp{\left( -\frac{k^2}{k^2_\textrm{fs}(t_\textrm{eq})} \right)} \qquad
\textrm{ if } k < k_\textrm{fs}(t_\textrm{eq}), \nonumber \\
& = & \exp{\left( -\frac{k^n}{k^n_\textrm{fs}(t_\textrm{eq})} \right)}\qquad
\textrm{   if } k \ge k_\textrm{fs}(t_\textrm{eq}),
\end{eqnarray}
with $n = 2$ at equality of radiation and matter \cite{Boyanovsky}.
At lower red-shift, $\tau^2(k)$ develops a non-linear regenerated
``tail" with $n$ \textit{measured} to be
between 0.5 and 0.8 at red-shift $z = 8$ \cite{UV} \cite{missing_satellite}.

In summary, limits on the ``thermal relic" warm dark matter mass
are not directly applicable to the \textit{colder} ``no freeze-in and no freeze-out" warm
dark matter that decouples early on from the Standard Model sector.
Three corrections would need to be applied:
\begin{enumerate}
\item
To correct for the different \textit{linear} warm dark matter free-streaming cut-offs,
due to their different temperatures,
the limit on $m_x$ needs to be multiplied by a factor $\le 0.30$,
depending on the details of the freeze-out and decoupling.
\item
An unknown correction needs to be applied due to the ``velocity
dispersion" cut-off. The ``velocity dispersion" delays and suppresses
the collapse of small scale density fluctuations, and defines the mass
of first galaxies, see Figure 6 of \cite{first}, and
see Table 1 of \cite{UV}. 
\item
An unknown correction needs to be applied due to the 
\textit{non-linear} regeneration of small scale structure
measured in \cite{UV} and \cite{missing_satellite}, that has contributions from baryons that are difficult
to include in hydro-dynamical simulations.
The measured $m_x$ is very sensitive to $n$, see \cite{comments}.
\end{enumerate}
Each of these items lowers the limit on $M_S$.

\section{Conclusions}

In conclusion, limits on the ``thermal relic" warm dark matter mass do not
apply to the \textit{colder} ``no freeze-in and no freeze-out" scenario of warm dark matter.
It is premature to dismiss several independent and consistent
\textit{measurements} of the warm dark matter temperature-to-mass ratio
summarized in \cite{poets}.

\end{document}